# Energy Level Sets for the Morse Potential

**Fariel Shafee**
Department of Physics
Princeton University
Princeton, NJ 08540

*Abstract:*

In continuation of our previous work investigating the possibility of the use of the Level Set Method in quantum control, we here present some numerical results for a Morse potential. We find that a proper treatment of the Morse potential eigenfunctions and eigenvalues for the case of a system with a small number of bound energy levels, the anharmonic perturbative approximation is actually invalid. We, therefore, use a Runge-Kutta integration method that gives more plausible results. We also calculate the dipole moment for the transitions between the two levels with our eigenfunctions and find that there is a critical depth of the potential. Finally we find the level sets giving equal expectation values of the energy, and comment on the unitary operators needed to make transitions from any level set to another.

## 1. Introduction

In a previous work [1] we have presented some general results regarding the quantum control of the expectation of an arbitrary observable, and in a later work [2] we have considered the application of the Level Set Method (LSM) [3], which is normally associated with classical fluid motion, in a quantum context. In both these works we were interested mostly in clarification and formal development of the relevant topics in terms of a general Hamiltonian.

In the present work we shall consider specifically the Morse potential, which is a potentially wonderful laboratory for studying quantum control [4], because it can give a finite number of bound states unlike a the Coulomb or the harmonic oscillator potential and hence the design of control is limited to a finite regime. At the same time the Morse potential is a good approximation to the experimentally observed vibrational modes of a diatomic molecule, and hence is not a toy model.

In the next section we shall first describe where and why the usual perturbative approximation used to calculate Morse eigenvalues may be erroneous. Since we shall concentrate on a two-bound -level system for our later considerations, we shall show numerically how variation of the parameters of the potential allows us to find such a system. We also calculate the dipole moments associated with the transitions between the two levels in terms of the potential depth. In section 3 we shall show how in a tri-atomic molecule we can get two-dimensional contours of equal expectation values of the vibrational energy, i.e. energy level sets. We shall then comment on the significance of two-level systems vis-à-vis quantum computing, and suggest methods for quantum control transcending usual quantum computing operations.

## 2. Morse Potential and its Energy Levels

The Morse potential may be described by many equivalent or nearly equivalent forms. We shall explicitly use the form (Fig. 1):

$$V(x) = c \ \{ \exp[-2\alpha(x-x_0)] - 2\exp[-\alpha(x-x_0)]\} \qquad (1)$$

where $c$ and $\alpha$ are parameters. Since we shall be working with an undefined length scale, we shall choose $x_0 = 1$ and $\alpha = 2$ arbitrarily.

Usually one can expand the potential above around its minimum at $x = x_0$, and since the first derivative is zero at $x_0$, the next term gives a simple harmonic approximation. The harmonic oscillator of course has an infinite number of equally spaced levels, which is unsuitable for quantum control considerations. If the quartic term is included we can make presumably make a better approximation, with energy levels given by

$$E_n = (n+\tfrac{1}{2})[1 - \alpha(n+\tfrac{1}{2})]\hbar\omega \tag{2}$$

where $n = 0, 1, 2, \ldots$, and $\alpha$ is a measure of the anharmonicity, i.e. the quartic term.

It is obvious that as $n$ increases the negative term overcomes the first term even if $\alpha$ is small. Hence it is assumed that the number of levels is finite.

However, this approximate relation is based on first order perturbation theory, and for perturbation theory to be valid, it is necessary to ensure that the series in convergent. First order correction involves $\langle n|x^4|n\rangle$ which can give ever increasing terms proportional to $n^2$ and that can soon overcome the smallness of $\alpha$. If we look at Fig. 1, we can see that the harmonic approximation fails miserably for large $x$, because the Morse potential asymptotically aligns with the x-axis, whereas the harmonic approximation would go on increasing. Hence, for large n, when the Morse potential allows a greater freedom of motion to large $x$, the HO potential confines it to much smaller values. So, for large $n$, Eq. 2 is not expected to be a good approximation although it gives nonlinearity of $E$ in terms of $n$ and a finite number of bound states.

Indeed, we should not expect to have any positive energy bound states when the potential approaches zero from below asymptotically, but Eq. 2 gives only positive energy solutions.

We have, therefore, used a coupled Runge-Kutta method to integrate Schrodinger equation from $x=0$ and used an iterative procedure to find the eigenvalue for different n values, which actually indicates the number of zeros of the eigenfunction. With $\alpha = 2$ we have found all possible bound states (different *n*'s giving negative *En*). Fig. 2 shows our results for *c* = *1* to *14*. We did not go beyond 14, because we are not interested in three-level systems for reasons we shall clarify later.

After the calculation of the energy eigenvalue we also find the eigenfunction. In Fig. 3 we show both the *n=0* and *n=1* eigenfunctions for the Morse potential with a typical value of *c*, viz. *c= 10*.

Using these eigenfunctions we have calculated the dipole moments:

$$d = <x> = \int dx \, \varphi_o(x) \, x \, \varphi_1(x) \tag{3}$$

It is very interesting to note [Table 1] that the dipole moment seems to depend critically on *c*, making a sudden transition of near zero values for *c< 7.2* to about *0.6* for *c* not smaller than *7.2*. Hence if we want to use laser radiation for quantum control we need a sufficiently deep Morse potential so that the action of the laser field can couple effectively with the dipole moment of the system, while it should not be so deep as to allow more than two energy levels if we want to restrict our considerations to the convenience of a two-level system.

## 3. Level Sets of Two-Potential Systems

Let us now consider a tri-atomic molecule with bonds between the central and the other two atoms only, described by two different Morse potentials. For example, it could be a HCN molecule. In case the bonds are nearly at right angles the vibrational modes may be described separately by the co-ordinates $x_1$ and $x_2$ of the extreme atoms and the state space can be simply a product space and the energy can be a sum.

$$|\Psi> \, = \, |\psi_1>|\psi_2>$$
$$E(n_1,n_2) = E_{n_1}+E_{n_2} \tag{4}$$

The separation is made more explicit if we describe the Lagrangian as

$$L = \tfrac{1}{2} m_1 \dot{x}_1^{\,2} - V(x_1) + \tfrac{1}{2} m_2 \dot{x}_2^{\,2} - V(x_2) \tag{5}$$

where $m_1$ and $m_2$ are the corresponding reduced masses.

For different end atoms we can have different parameters for the two potentials and also different masses, giving different energy eigenvalues and eigenfunctions. If the value of $c$ is still within the values we have discussed above, then we shall have two energy levels for both sectors.

With only two–energy levels in each sector, because of the normalization condition a mixed state in each sector can be described by only one real (bound states are described by real functions) constant, or equivalently, by an angle of rotation in the two-dimensional state space.

Using the superscript for the atom and the coefficients for the corresponding $|0\rangle$ and $|1\rangle$ states:

$$|\psi_i\rangle = a_i^0 \, |0_i\rangle + a_i^1 \, |1_i\rangle \tag{6}$$

we get, as shown in ref [2], ellipses in the $(a^0_1, a^0_2)$ plane:

$$(a_1^0)^2 (E_1^1 - E_1^0) + (a_2^0)^2 (E_2^1 - E_2^0) = E_1^1 + E_2^1 - \langle E \rangle \tag{7}$$

the left and right hand side terms being always positive. As $\langle E \rangle$ changes, the system moves from one ellipse to another with modified major and minor axes, contracting to the origin when the $\langle E \rangle$ value becomes equal to the maximal possible value $E^1_1 + E^1_2$, as expected.

We have the constraints that the coefficients cannot exceed one in magnitude (limiting square in Fig. 4). The entire ellipse can become a valid level set only if both the following conditions are met:

$$\begin{aligned} E_1^0 + E_2^1 \leq \langle E \rangle \quad & E_1^0 + E_2^1 \leq \langle E \rangle \\ E_1^1 + E_2^0 \leq \langle E \rangle \quad & E_1^1 + E_2^0 \leq \langle E \rangle \end{aligned} \tag{8}$$

In Fig. 4 we show two level sets with the coefficient $a^0_1$ for a potential with $c = 10$ along x-axis, and the coefficient $a^0_2$ for a potential with $c = 12$ along the y-axis.

## 4. Quantum Transitions between Level Sets

As we have mentioned in ref [2], unlike classical transitions between level sets, in the quantum context the transitions may be dictated by convenience from among virtually an infinite number of paths available. The unitary operator will be actually to independent operators in the case we have here discussed.

$$U = U_1 \otimes U_2 \ U = U_1 \otimes U_2 \tag{9}$$

Given any initial state:

$$|\Psi> = \begin{pmatrix} \cos\theta_1 \\ \sin\theta_1 \end{pmatrix} \otimes \begin{pmatrix} \cos\theta_2 \\ \sin\theta_2 \end{pmatrix} \tag{10}$$

we can go to any final state with a similar expression but different $\theta_1$ and $\theta_2$ simply by a unitary operator representing the appropriate rotations in the two spaces. This can be achieved by laser irradiation, as shown by Ramakrishna et al. [4]. The interesting point is that we can have two different laser pulses with different frequencies in sequence, corresponding to the two spaces, if they have different energy eigenvalues, and we can also use our knowledge of the dipole moments which we have calculated to fix the duration and amplitude of the laser pulses for optimal performance as explained in the reference mentioned above.

## 5. Conclusions

Two-level Morse potentials offer the possibility of being treated as qubits of quantum computing. One can then in principle form quantum registers with arrays of such molecules in close proximity offering the possibility of entanglement. It is not clear yet what kind of mechanisms can serve as Hadamard, c-NOT and other gates to provide full computing facilities. To use the Grover [5] algorithm, for example, one needs a phase inversion mechanism, which we have not considered here. But, again in principle, one can probably always use laser beams to perform any unitary operation, and that would include a phase inversion.

The reverse scenario is also interesting. In a previous work [1] we have indicated that we must know the present state of a system to take it to a target state. However, that information may not be necessary if we know that the initial state contains a given state as a component. We can then use a sequence of unitary operations as in the Grover algorithm to take it arbitrarily close to an intermediate state by a finite number of unitary operations and therefrom to the target state. This opens up the possibility of making transitions from a given level set to another by a universal set of operations without specific knowledge of the exact states involved.

I would like to thank Professor H. Rabitz for reading the manuscript and for discussions, and Andrew Tan of UCSF for encouragement.

**Table** I: dipole moment (d) for various depths (c) of the potential-

| c | 6       | 7       | 8      | 9      | 10     | 11      | 12     | 13     | 14     |
|---|---------|---------|--------|--------|--------|---------|--------|--------|--------|
| d | -0.0003 | -0.0003 | -0.588 | -0.604 | -0.609 | -0.6089 | -0.607 | -0.602 | -0.597 |

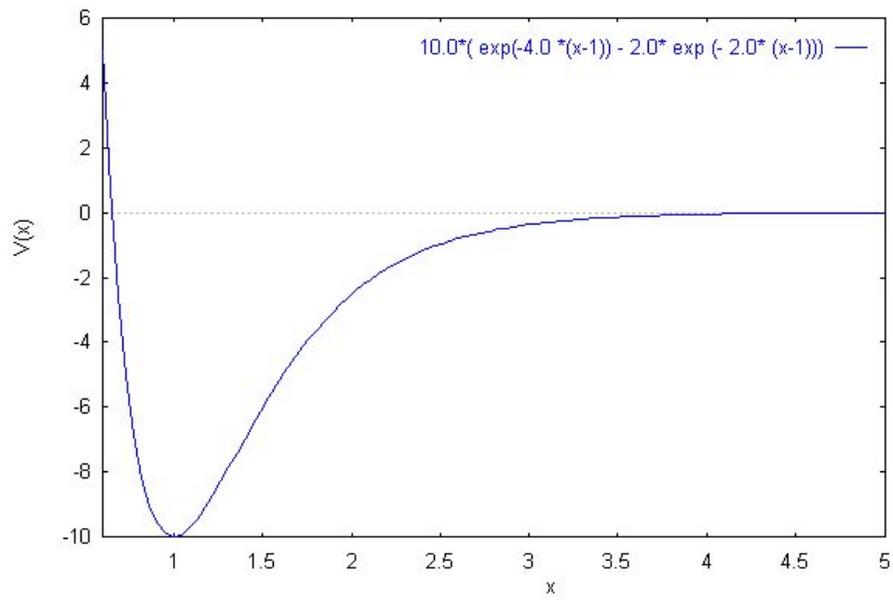

Fig.1: Morse potential: V(x) = 10 { exp[-4(x-1)] – 2 exp[-2 (x-1)] }

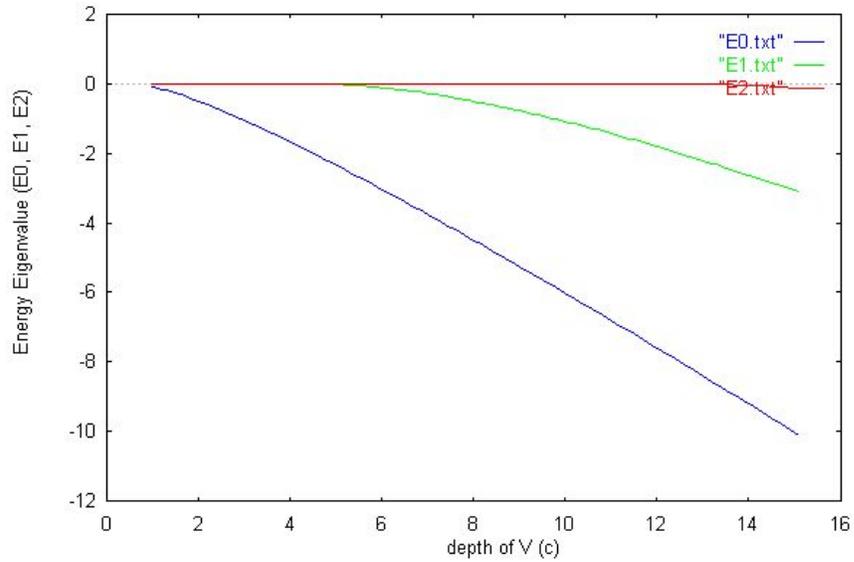

Fig. 2: Energy eigenvalues of the first three states for the Morse potential as a function of the potential depth.

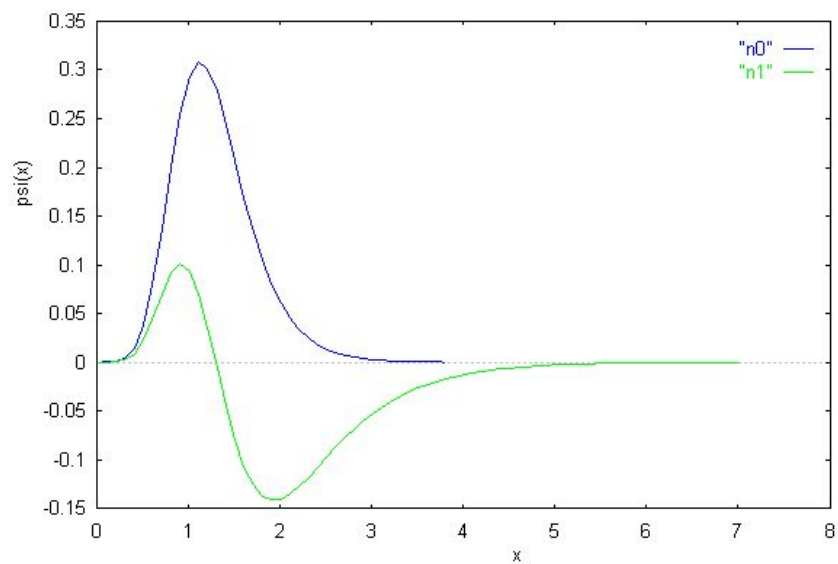

Fig. 3: Eigenfunctions for *n = 0* and *n = 1*.

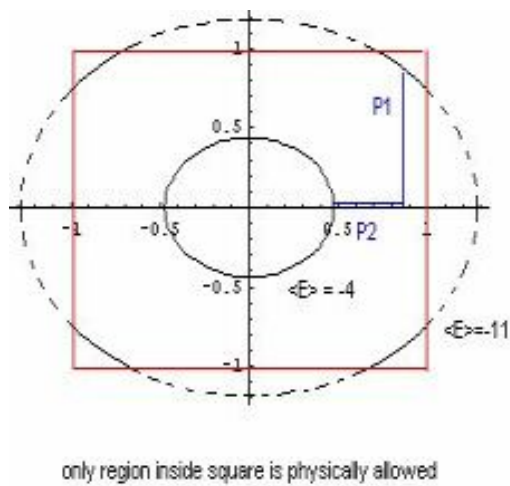

only region inside square is physically allowed

Fig.4: Two level sets corresponding to <E> = -4 and –11. P1 and P2 are possible pulses to make a coherent transition from outer level set to inner using different frequencies for the two different potentials.